\documentclass[aps, prl, showpacs, twocolumn, amsfonts, amsmath, amssymb, floatfix, groupedaddress,longbibliography]{revtex4}

\usepackage{graphicx}
\usepackage{dcolumn}
\usepackage{bm}
\usepackage{hyperref}
\usepackage{amsmath}
\usepackage{xcolor} 

\begin{document}

\title{Giant diffusion of nanomechanical rotors in a tilted washboard potential}

\author{L. Bellando$^1$, M. Kleine$^1$, Y. Amarouchene$^1$, M. Perrin$^1$, Y. Louyer$^1$}

\affiliation{$^{1}$Universit\'{e} de Bordeaux, CNRS, LOMA, UMR 5798, F-33405 Talence, France}

\date{\today}

\begin{abstract}
We present an experimental realization of a biased optical periodic potential in the low friction limit. The noise-induced bistability between locked (torsional) and running (spinning) states in the rotational motion of a nanodumbbell is driven by an elliptically polarized light beam tilting the angular potential. By varying the gas pressure around the point of maximum intermittency, the rotational effective diffusion coefficient increases by more than 3 orders of magnitude over free-space diffusion. These experimental results are in agreement with a simple two-state model that is derived from the Langevin equation through using timescale separation. Our work provides a new experimental platform to study the weak thermal noise limit for diffusion in this system.
\end{abstract}


\maketitle

\begin{figure}[h!]
\centering
\begin{minipage}[b]{9cm}
\includegraphics[width=9cm]{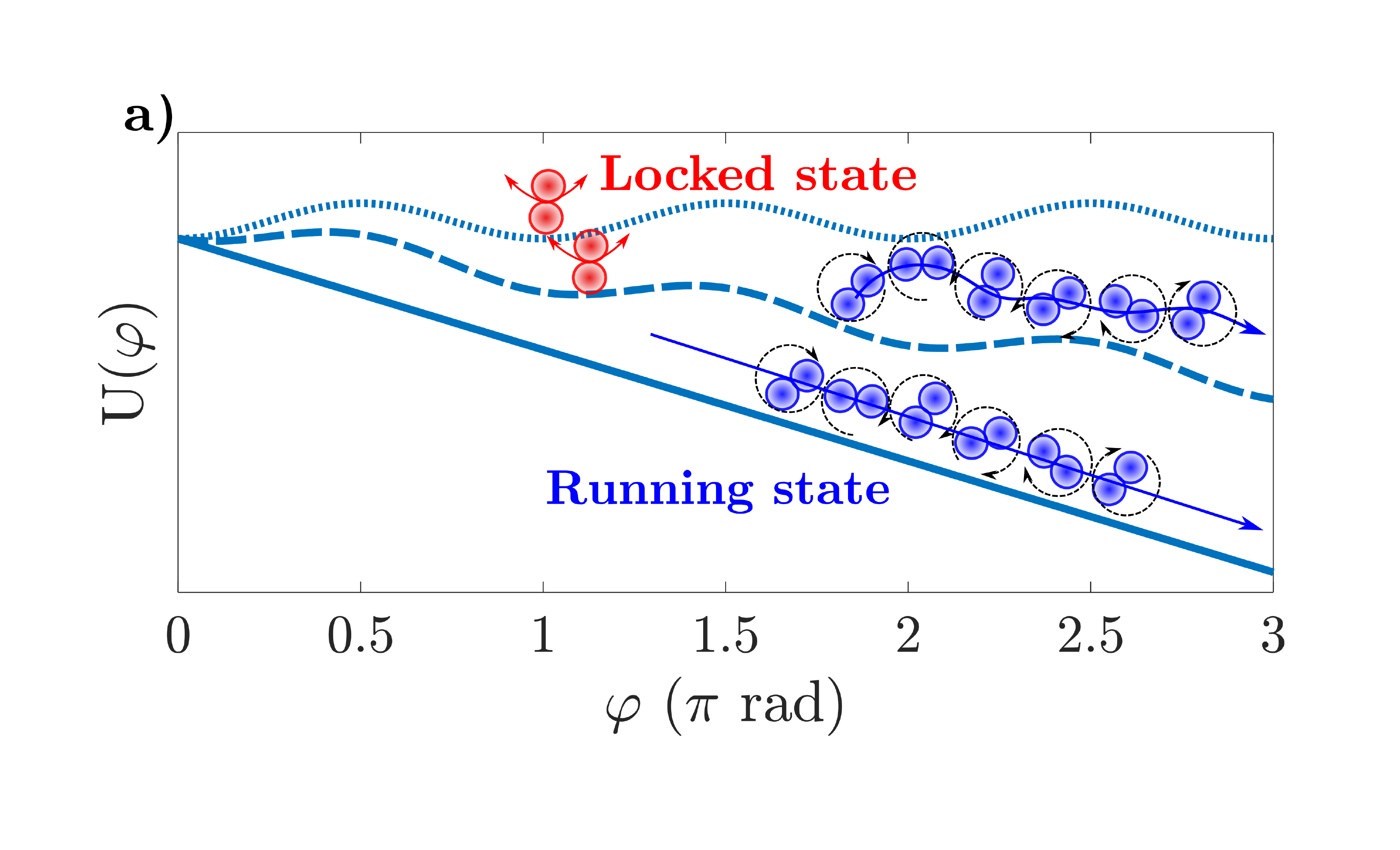}
\end{minipage}
\begin{minipage}[b]{9cm}
\includegraphics[width=9cm]{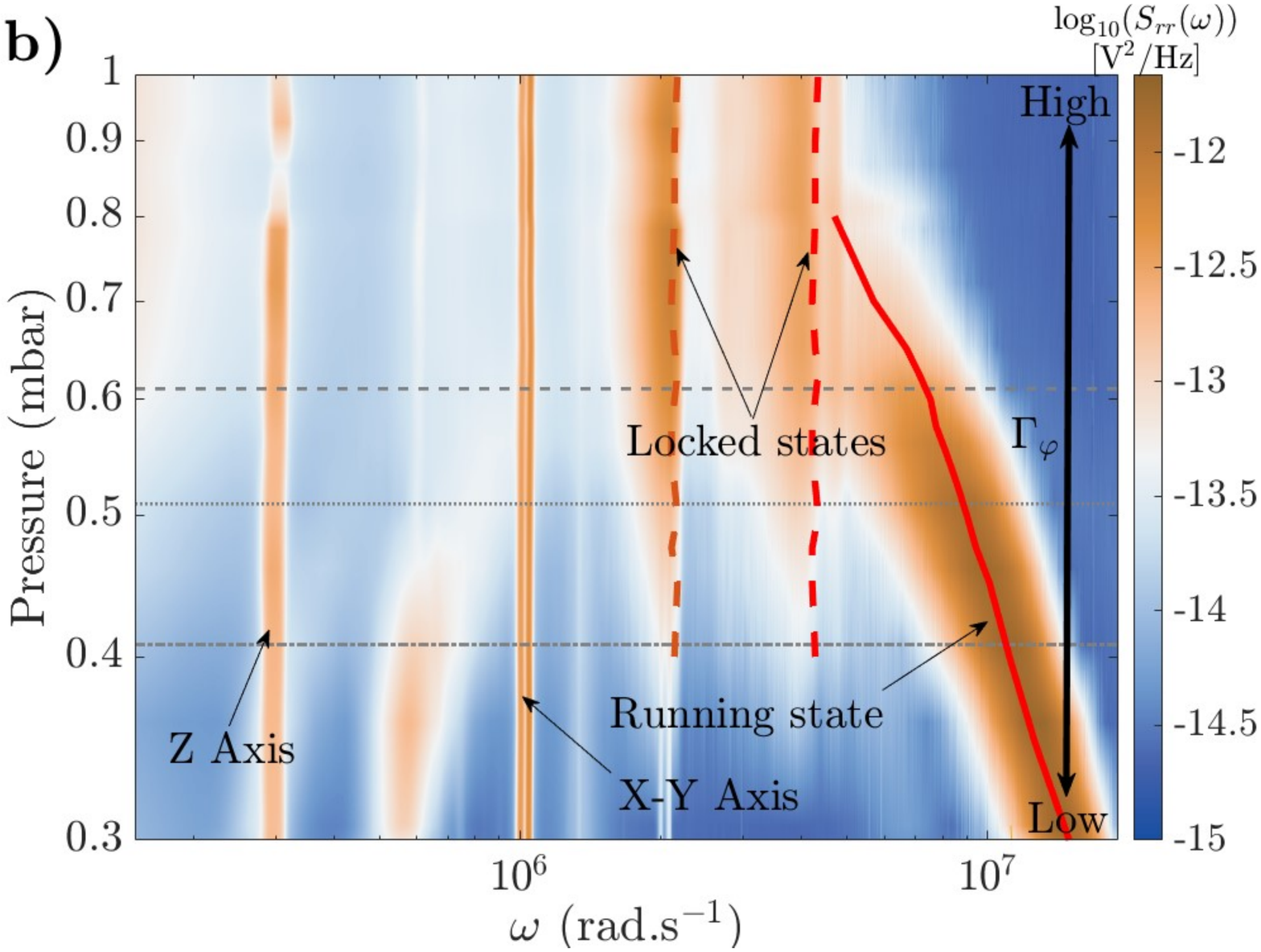}
\end{minipage}
\caption{ \textbf{a)} Illustration of the energy landscape for the rotation of the Brownian dimer around the optical axis for three different tilts. When the polarization is linear (blue dotted line), the potential is not tilted and the orientation of the dimer remains in the locked state. Conversely, for a circular polarization (solid blue line) the potential wells disappear. For elliptical polarization (dashed line), the rotational motion can change from a running state (spinning) to a locked state (torsional) and vice versa. \textbf{b)} Surface plot in log-log scales of the angular vibration and translational power spectral densities as a function of the gas pressure for a beam ellipticity of $\phi_{\lambda/4}=25^{\circ}$ and an optical power of $180$~mW. The solid and dashed lines (red) represent the local maxima of the PSD computed by Langevin simulations of Eq.~(\ref{LangR}). The three grey horizontal lines at 0.41 mbar, 0.51 mbar and 0.61 mbar outline the two-state coexistence region where the middle line denotes the maximum of intermittency.}
\label{fig1}
\end{figure}

\begin{figure}[h!]
\centering
\begin{minipage}[b]{9cm}
\includegraphics[width=9cm]{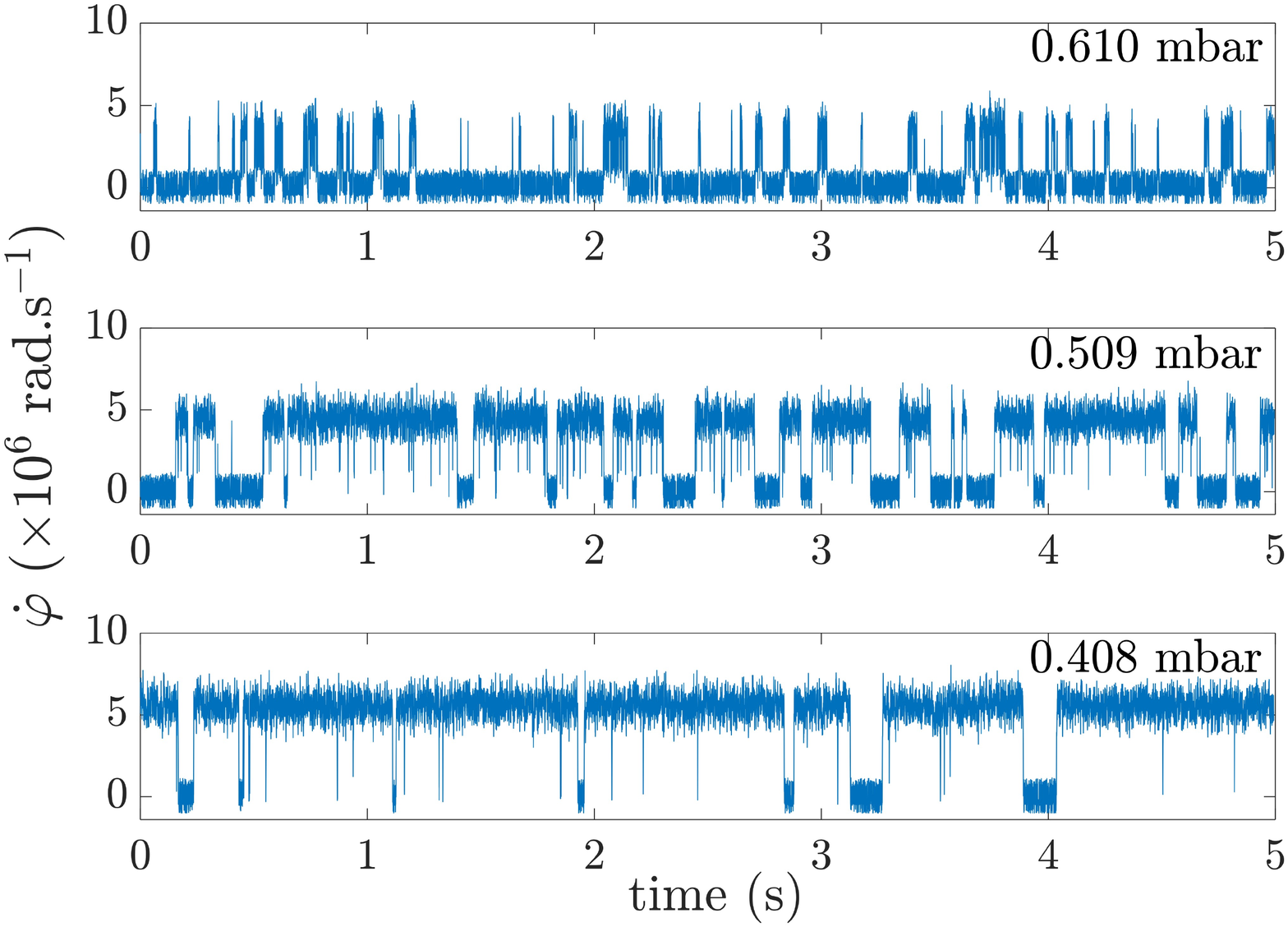}
\end{minipage}
\caption{  Time traces of the dimer angular velocity $\dot{\varphi}$ obtained by short-time Fourier transform for the three pressures that are shown as gray horizontal lines in Fig.~\ref{fig1}(b).}
\label{fig2}
\end{figure}

Thermal diffusion in a tilted periodic (also named washboard) potential constitutes an archetypal model of transport phenomena in nonequilibrium statistical physics \citep{KRAMERS1940284,risken1989fpe}. It describes a diverse range of systems, including Josephson junctions \citep{JosephsonPRB}, molecular motors \cite{Hayashi15}, synchronization phenomena \cite{Strogatz03}, and diffusion on crystal surfaces \cite{Frenken85}, to name only a few.

Brownian particles moving in the tilted potential at non-zero temperature exhibit two distinct well-characterized limiting behaviors: locked and running states. The first emerges when the potential wells are deeper than the thermal energy scale, the diffusing particle remains in a local minimum. In the latter, a sufficient tilt of the potential, lowering its energy barrier, allows the particle to flow down the potential wells. Depending on the system parameters (friction, temperature and tilt), both solutions may coexist, and stochastic transitions between locked and running states may occur. It has been shown, both experimentally \cite{GrierPRL,Hayashi15,Evstigneev08} and theoretically \citep{PRL2001_Reimann, Costantini_1999}, that in this two-state coexistence regime, the effective diffusion coefficient can be enhanced by several orders of magnitude relative to the free diffusion coefficient. However, this has only been experimentally observed in the overdamped regime which is by now a fully understood stochastic process \citep{risken1989fpe}.

To our knowledge, an experimental study of giant diffusion in the strongly inertial (underdamped) regime is still missing. The theoretical problem is also more demanding in the underdamped and weak noise limits because the usual expansion of the matrix continued fraction \citep{risken1989fpe} fails to converge well \citep{SokolovPRE}. On these grounds, this nonlinear stochastic system in the low-noise and low-damping limits is still attractive \citep{SokolovPRE}.

We propose a simple experimental setup based on the optical trapping of a nanodumbbell in a moderate vacuum. A special feature of our experiment is that the friction coefficient, being linearly proportional to the gas pressure $P$, can be tuned over several orders of magnitude. Therefore, for a nearly constant temperature, Brownian motion can move from the overdamped to the underdamped regime. This makes the optical trapping setup a suitable platform to study stochastic dynamics in the low-friction regime \cite{Amarouchene19,Mangeat19,Rondin17}.

The main focus of this Letter is the analysis of the rotational diffusivity in the bistability region for an underdamped stochastic mechanical nanorotor travelling in a tilted periodic potential. We provide experimental confirmation of a giant diffusivity by tracking the rotational motion of a single silica nanodimer. Due to the experimentally achieved timescale separation, we show that the effective diffusion is well described by a two-state model for this nearly one-dimensional rotational motion.

The experimental setup consists of a vacuum optical tweezer  trapping a dielectric silica dumbbell made of two nanospheres of nominal radius $68\pm7$~nm \citep{TongGHzDumbell,NovGHzSphere}. The continuous wave trapping laser beam (wavelength, $\lambda=1064~$nm and power, $180~$mW) passes through an objective lens with a $0.8$ numerical aperture, the polarization's ellipticity of which can be changed by a quarter-wave plate. A detailed description of the experimental setup, calibration and processing procedures is given in the Supplemental Information \citep{SI}. The shape and the size of the dimer are assessed under linear polarization, and correspond to an aspect ratio $L/D=1.8,$  with two spheres of radius $R_p=65.7~$nm, where $D=2R_p$ and $L$ is the major axis length of the dimer.
Our detection system is able to simultaneaously record the particle's center of mass motion and its angular displacement.

The nanodimer experiences both a trapping force and a torque, both of which are induced by elliptically polarized light. In particular, when such an asymmetric Rayleigh scatterer lies in the transverse ($\hat x, \hat y$) plane, corresponding to $\theta=\pi/2$, the torque reads
\begin{equation}
M=M_{C}-M_{L}\sin(2\varphi),
\label{Tau_Def}
\end{equation}
where $\varphi$ is the angle between the dimer main axis and the lab's horizontal $\hat x$ axis. Its time evolution is related to the dimer rotation around the $\hat z$ optical axis. $M_C$ and $M_L$ are respectively the \textit{circular} and \textit{linear} contributions of the optical torque which depend on the dimer polarizability tensor and the laser beam polarization \citep{SI}.

In a linearly polarized optical tweezer, where $M_C=0$, the long axis of the dimer (\textit{i.e.,} in the direction of greatest polarizability) will tend to align with the laser beam polarization. This is because the polarizability of the dimer along its long axis is greater than the polarizability perpendicular to it.
As a result, the dimer acts as a torsion balance with a linear restoring torque, $-2M_{ L}\varphi$.  Note that regardless of the gas pressure, the dimer oscillates about $\varphi=0$ with a torsional frequency $\Omega_{\rm L}=\sqrt{2M_{L}/I_\perp}$, where $I_\perp$ is the moment of inertia.
In contrast, for a circularly polarized trap, where $M_L=0$, the asymmetric particle undergoes a constant torque, whatever its orientation $\varphi$. As a result of the continuous transfer of angular momentum, the particle will spin about the $\hat z$ axis at constant frequency, as given by $\Omega_{\rm R}=M_{C}/I_\perp\Gamma_{\varphi}$, where $\Gamma_{\varphi}$ is the rotation or torsional vibration damping rate about the $\hat z$ optical axis (see \citep{SI} for its determination). Note that the rotation frequency can exceed GHz, which allows us to study material stresses due to centrifugal forces \cite{AshkinAPL,OriolUltimTensilStrength}, measure ultraweak torques \cite{DholakiaReviewRotation, UlbrichtPrecessionPRL}, and examine quantum features of rotation \cite{QuantumRotationNanoPart,Stickler18}.
Figure~1(a) illustrates the potential landscape and orientational dynamics for different polarization states. Considering the general case, for elliptical polarization, the potential has a washboard shape:
$U(\varphi)=M_{L}\sin^2(\varphi)-M_{C}\varphi$, and one can observe running and locked states. Below, we only consider elliptical polarization which is held constant at $\phi_{\lambda/4}=25^{\circ}$, the angle of the quarter waveplate.


The dynamics of an asymmetric Brownian particle trapped under elliptically polarized light is 6-dimensional with translational and rotational degrees of freedom coupled to each other. This generally leads to a highly nonlinear problem \cite{UlbrichtPrecessionPRL,LiNLHybrid,DholakiaNatComRotCool}. However, in the following, we show that the giant diffusion phenomenon is well described by one-dimensional rotational dynamics. Comparison between the experiment and numerical simulations is made quantitative provided that an additional term is added to the usual \textit{extinction} optical torque ($\propto \Re[\textbf{p}\times \textbf{E}_\textbf{\rm inj}]$, with $\textbf{p}$ the dipole moment and $\textbf{E}_\textbf{\rm inj}$ the incident electric field). This extra term, called \textit{scattering} torque ($\propto \Im[\textbf{p}\times \textbf{p*}]$), is due to the interference of the fields scattered by the particle \citep{Nieto_Vesperinas_Torque,SI}.
This $\varphi$ angle dynamics reproducing the stochastic jumps between torsion and continuous rotation states is described by the Langevin equation
\begin{equation}
I_\perp\ddot{\varphi}=-I_\perp\Gamma_{\varphi}\dot{\varphi}+M + M_{\rm th},
\label{LangR}
\end{equation}
where $\Gamma_{\varphi}$ is the rotational damping rate and $M_{\rm th}=\sqrt{2k_BTI_\perp\Gamma_{\varphi}}\zeta(t)$ is the thermal torque with $\zeta(t)$ a gaussian noise of zero mean value. $M$ is given by Eq.~(\ref{Tau_Def}).

Figure 1(b) shows the position and orientation power spectral densities as a function of pressure, along with the numerical Langevin simulations (solid red lines). For consistency, we also show the frequencies associated to translational motions in the $X,Y$ and $Z$ directions that are independent of gas pressure. The observed non-degeneracy of the transverse frequencies $X$, $Y$ is expected for a non-circularly polarized laser beam.
In the high friction limit, the torsional dynamics dominate the dimer orientation motion and their resonance frequencies remain constant in relation to pressure. Conversely, the rotational dynamics is dominant in the low friction regime, i.e., at low pressure, and its angular frequency, $\Omega_{\rm R}$, rises as the inverse of the gas damping.

The red solid lines in Fig.~1(b) show a good agreement between the experiment and one-dimensional simulations of Eq.~(\ref{LangR}). One can also clearly see that the bistability region is quantitatively well-described by the model. The two locked states observed in the spectral densities are also found in the numerical simulations, and correspond to the fundamental peak and its overtone.
Although not within the scope of this study, we do not preclude contributions with other angles leading to coupling with extra motional degrees of freedom. The experimental determination of the dimer size and shape by the calibration procedure \citep{SI} is used for calculating the moments of inertia and polarizabilities. The latter are further supported numerically using Comsol Multiphysics. We find that $I_\perp = 2\times 10^{-32}$~kg$\cdot$m$^2$, $M_C$ and $M_L$ are respectively of the order of $2\times 10^{-22}$~N$\cdot$m and $6\times 10^{-20}$~N$\cdot$m, in good agreement with both experimental and numerical results.

As mentioned above, in the calculation of optical torques the contribution of \textit{scattering} is included in addition to the usual \textit{extinction} contribution \cite{Nieto_Vesperinas_Torque}.
Since the polarization of the trapping beam is not linear, this scattering torque is crucial to reproduce quantitatively the experimental results, particularly when the dimer spins. In many previous works, the scattering torque was ignored because the rotational Langevin equation was only qualitatively discussed.  It is worth noting that if the scattering torque is not taken into account, the particle's rotation speed is overestimated by typically an order of magnitude for the actual dimer \cite{SI} and much more for a spheroid (data not shown).
Some quantitative deviations between theory and experiment appear at low pressure but without greatly affecting the giant diffusion effect.

\begin{figure}[h!]
\begin{minipage}[b]{9cm}
\hspace{-0.8cm}
\includegraphics[width=8.5cm]{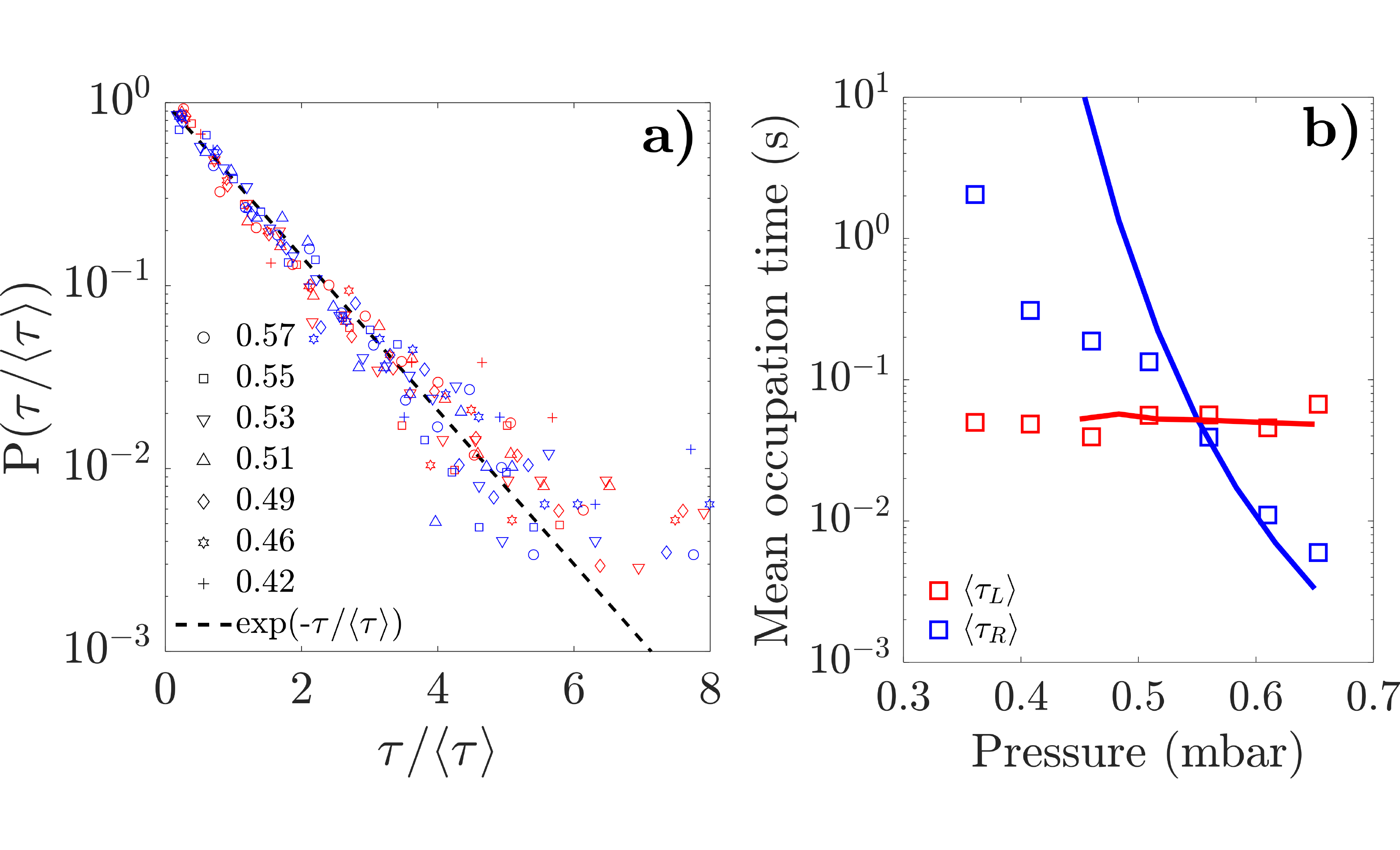}
\end{minipage}

\caption{\textbf{a)} Probability distribution function of the occupation time from the locked to running states $P(\tau_{L})$, red symbols, and the running to locked states $P(\tau_{R})$, blue symbols, for various pressures where the bistability  is observed. Exponential behaviors suggest that stochastic transitions between locked and running states are thermally activated, following an Arrhenius-type law. \textbf{b)} Pressure dependence of the mean occupation times in the locked and running states [symbols: experiment, solid line: Langevin simulations, Eq.~(\ref{LangR})].  }
 \label{fig3}
\end{figure}

We now consider the quantitative study of this stochastic nonequilibrium phenomenon. Fig.~2 shows the time traces of the angular velocity $\dot{\varphi}$ obtained by short time Fourier transforms in the bistability region. In the locked state, the mean angular velocity is $\langle\dot{\varphi}\rangle_L \equiv0$. While the mean rotational speed, $\langle\dot{\varphi}\rangle_R$, in the running state increases continuously with the pressure drop. The most noticeable feature of the time traces is the two-state noise-induced transitions, where the maximum of intermittency occurs in the middle of the bistability region at about $P \approx 0.56$~mbar [Fig. 1(b)].

\begin{figure}[h!]
\centering
\begin{minipage}[b]{8.5cm}
\includegraphics[width=8.5cm]{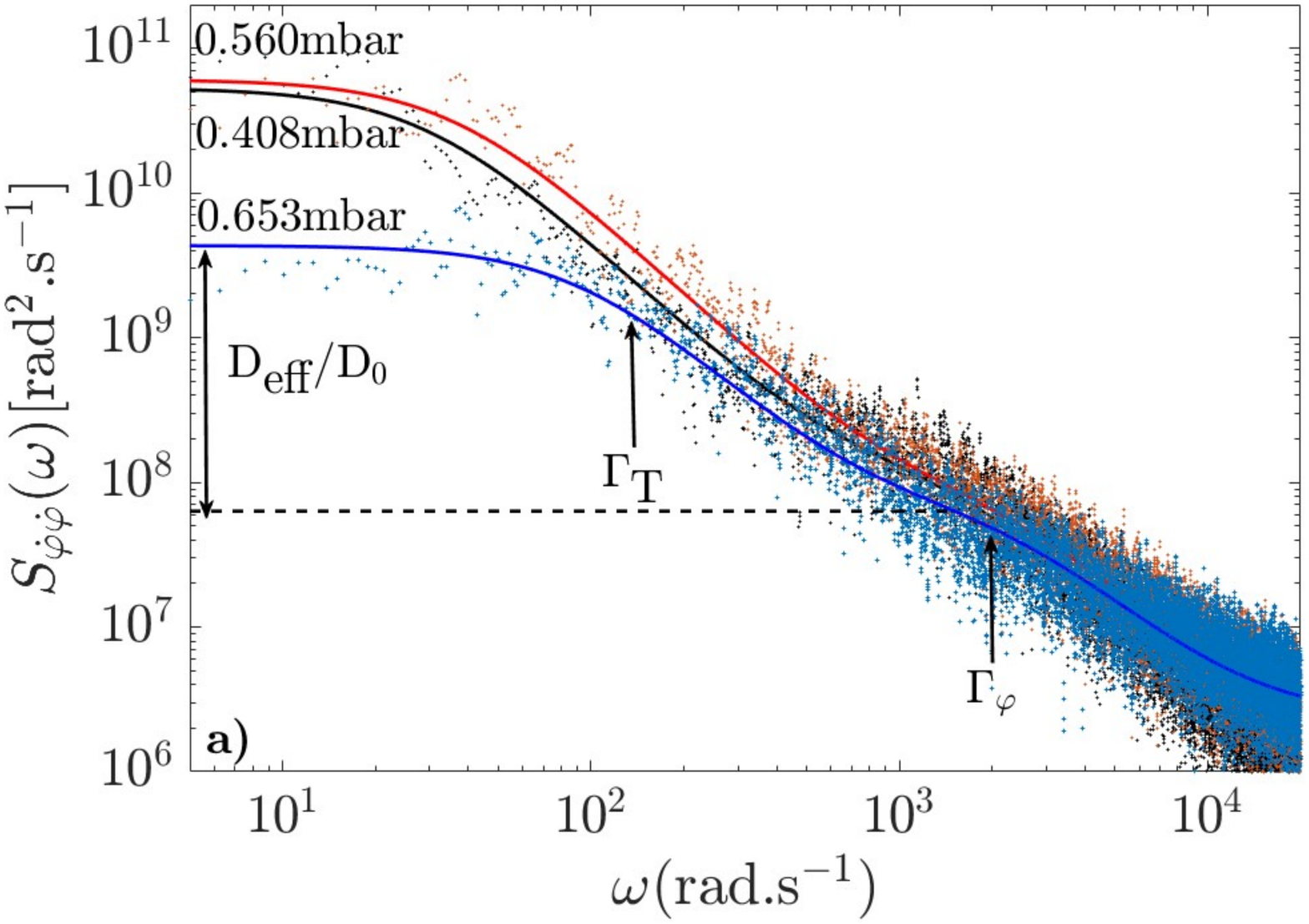}
\end{minipage}
\vspace{0.1cm}
\begin{minipage}[b]{8.5cm}
\includegraphics[width=8.5cm]{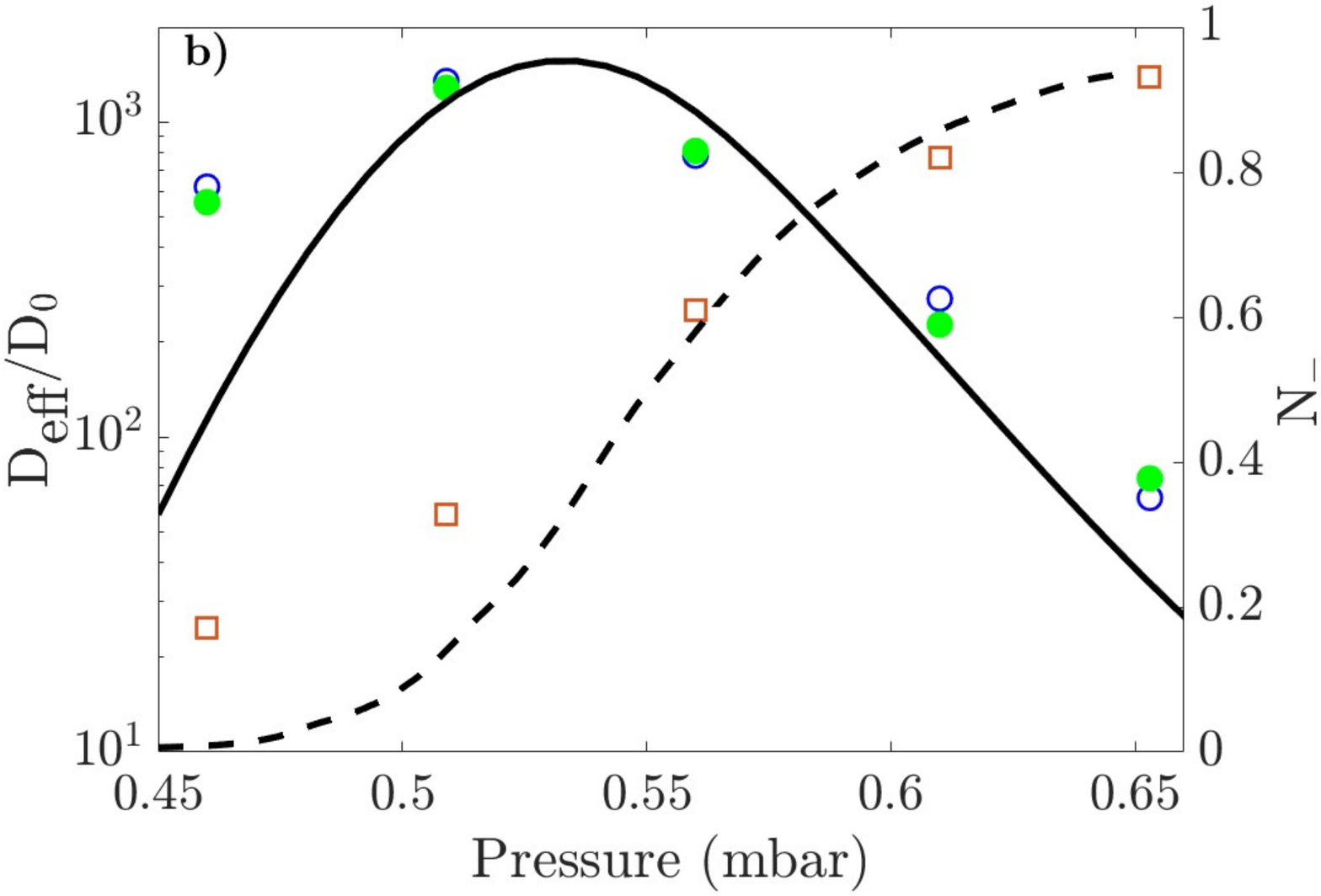}
\end{minipage}
\caption{\textbf{a)} Measured power spectral densities of the dimer angular velocity (dots) and double Lorentzian fit (Eq.~(\ref{SOmOm}), solid lines) at the boundaries of the bistable region ($P=0.653$~mbar in blue, $P=0.408$~mbar in black), and at the pressure around which the maximum of giant diffusion exists ($P=0.560$~mbar in red). The dashed line represents the free-space diffusion coefficient $D_0$ at $P=0.653$~mbar, the double arrow displays the excess of diffusion $D_{\text{eff}}/D_{0}$, while both the total escape rate $\Gamma_{\rm T}$ and the rotational damping rate $\Gamma_{\varphi}$ are marked by vertical arrows. \textbf{b)} On the left axis. Pressure dependence of the effective diffusion coefficient $D_{\text{eff}}$ in units of $D_0$ estimated by two methods: (i) data from Fig. 4(a) are fitted by using Eq.~(\ref{SOmOm}), blue open circles, while (ii) transition rates obtained from Fig.~(3) are used to calculate $D_{\text{eff}}$ from Eq.~(\ref{Deff}), green filled circles. The black solid line results from the Langevin simulations [Eq.~(\ref{LangR})]. On the right axis. Proportion of the mean occupancy in the locked state (red squares: experiment and black dashed line: Langevin simulations).}
 \label{fig4}
\end{figure}

We use the time traces of angular velocities to determine the mean occupation times in the locked and running states. Fig.~3(a) shows the distributions $P(\tau/\langle\tau\rangle)$ of the mean occupation times, recorded for different gas pressures. These distributions decrease exponentially as expected for a Kramers-like problem. Deeper insight into two-state coexistence can be gained if we plot the pressure dependence of the average occupancy times in the locked and running states, as shown in Fig.~3(b), both for experimental and numerical data. It can be evidenced that when $\Gamma_\varphi \ll \Omega_L$ (as in this case), the transition rate $r_L$ out of a locked state (regardless of the final state, locked or running) is independent of the friction coefficient, i.e., of the pressure. This is in agreement with our observations (see Fig. 3(b) red squares and solid line). Besides, this value is well approximated by the well-known Arrhenius law, $1/\tau_L=r_L \sim \Omega_L \exp(-E_{\rm b}/k_{\rm B}T),$ $E_{\rm b}$ being the barrier height given by the potential difference between a minimum and the lower neighboring maximum \cite{LandauerPRB}. This offers insight in how a temperature change affects the stability of the torque state. A different behavior is observed for transitions from a running to a locked state for low damping and finite temperatures. This behaviour is explained by the fact that the rate $r_R$ is such that $1/\tau_R=r_R \sim \Gamma_\varphi \propto P$ \cite{Buttiker83}. Note that some differences occurring at low pressure can be observed betwee*n the one-dimensional model and the experiment [as seen in Fig. 3(b)]. They are not due to a time windowing effect, but could arise from the coupling between other rotational degrees of freedom, as seen below 0.45~mbar with the appearance of an additional resonance in the spectral densities [Fig.~1(b)].
In the rest of this Letter, we focus on the measurement of the effective rotational diffusion coefficient, which is defined by the Kubo relation
\[D_{\rm eff} = \int^\infty_0 dt' \langle [ \dot{\varphi}(t') - \langle \dot{\varphi} \rangle][ \dot{\varphi}(0) - \langle \dot{\varphi} \rangle] \rangle.\]
 This coefficient satisfies the usual Einstein relation $\lim_{t \rightarrow \infty} \langle \varphi^2(t) \rangle - \langle \varphi(t) \rangle^2 = 2 D_{\rm eff} t$, where $D_{\rm eff}$ can in principle be enhanced by orders of magnitude over the free coefficient diffusion $D_0$ at the crossover from locked to running states. Interestingly, outside the region of bistability, the power spectral density of $ \dot{\varphi}$ allows for a measurement of both $D_0=k_{\rm B}T/\Gamma_\varphi I_\perp$ and its related timescale $1/\Gamma_\varphi$. Surprisingly, we can derive a simple two-state stochastic noise model to fit our experimental results in the bistability region (see details in Sec. V.B of the Supplemental Material \cite{SI}). Since the parameter range we consider allows large timescale separations ($1/\Omega_{\rm R} \ll 1/\Gamma_\varphi \ll {\tau_L, \tau_R}$), the spectral density of the rotation speed in the region of bistability reads
\begin{equation}
S_{\dot{\varphi}\dot{\varphi}}(\omega)=\frac{D_0}{1+\frac{\omega^2}{\Gamma_{\varphi}^2}}+\frac{D_{\text{eff}}}{1+\frac{\omega^2}{\Gamma_{T}^2}}
\label{SOmOm}
\end{equation}
where $\Gamma_{\rm T} = r_R + r_L,$ is the total escape rate, corresponding to the longest timescale of the system. From the time traces of the angular velocity $\dot{\varphi}$ (Fig.~2), we compute the power spectral densities of the rotation velocity (defined as the Fourier transform of the rotational velocity autocorrelation function) in the coexistence regime, as shown in Fig.~4(a). Using then Eq.~(\ref{SOmOm}), we can determine the effective diffusion $D_{\rm eff}$ relative to $D_0$, that is displayed in Fig.~4(b) as a function of pressure. Note that $\Gamma_\varphi$ and $\Gamma_{\rm T}$ are also measured, allowing to corroborate the values of the nanodimer size and its aspect ratio. To go further into the two-state model, we use the transition rates $ r_R $ and $ r_L $ deduced from Fig.~3 to calculate $D_{\rm eff}$. In our parameter regime, the  two-state model gives an expression that bears much similarities with those developed in \citep{SokolovPRE,OriginalShotNoise}
\begin{equation}
D_{\text{eff}}=\langle\dot{\varphi}\rangle_{_{\rm R}}^2 ~\frac{r_{L}r_{R}}{(r_{L}+r_{R})^3},
\label{Deff}
\end{equation}
where $\langle\dot{\varphi}\rangle_{_{\rm R}}$ is the average angular velocity taken only over the running states of the whole temporal trace. Comparison of the filled and open circles in Fig.~4(b) indicates that when Eq.~(\ref{Deff}) is used with the results obtained from occupancy times, good agreement is obtained with the diffusion estimate by spectral densities by using Eq.~(\ref{SOmOm}).
The occupancy rate in the locked state defined as $N_-=\langle\tau_{L}\rangle/(\langle\tau_{L}\rangle+\langle\tau_{R}\rangle)$ is represented by red squares in Fig.~4(b). We observe that the giant increase in the effective diffusion coefficient is roughly maximal at a pressure where the states are evenly distributed.

In conclusion, we have observed in a very small pressure range a giant increase of the diffusion coefficient in the underdamped and weak noise limit. Carefully calculating the optical torque, a reasonable quantitative agreement has been obtained using a one-dimensional Langevin model. This supports evidence that even though the real system dynamics has 6 dimensions, it can be reduced to a one-dimensional system within at least some parameter regimes.
Cooling the translational (and possibly rotational) degrees of freedom would mitigate the rotranslational coupling \citep{NovGHzSphere,TongGHzDumbell}, enabling ideally over a wide parameter regime to reach the true one-dimensional Brownian nanorotor as described in Eq. (2). Thus, an optically levitated stator-rotator constitutes an outstanding platform where its dynamics can be controlled in many ways up to the quantum ground state at room or cryogenic temperature \citep{Delic20,AspelmayerNat,NovotnyCryoNat}.

The authors acknowledge financial support from the Region Nouvelle-Aquitaine (grant 2018-1R50304) and the Agence Nationale pour la Recherche (grant ANR-21-CE30-0006). M.~K. acknowledges the financial support from the EUR Light S\&T Graduate Program (PIA3 Program “Investment for the Future”, ANR-17-EURE-0027).

\bibliography{MyBibGiantDiff_4}

\end{document}